# UCN sources at external beams of thermal neutrons. An example of PIK reactor.

E.V. Lychagin[1], V.A. Mityukhlyaev[2], A.Yu. Muzychka[1], G.V. Nekhaev[1], V.V. Nesvizhevsky[3], M.S. Onegin[2], E.I. Sharapov[1], and A.V. Strelkov[1]

[1] *Joint Institute for Nuclear Research, Dubna, Russia,*
[2] *Petersburg Nuclear Physics Institute, Gatchina, Russia,*
[3] *Institut Max von Laue – Paul Langevin, Grenoble, France.*

**Abstract**

We consider ultracold neutron (UCN) sources based on a new method of UCN production in superfluid helium ($^4$He). The PIK reactor is chosen as a perspective example of the application of this idea, which consists of installing a $^4$He UCN source in a beam of thermal or cold neutrons and surrounding the source with a moderator-reflector, which plays the role of a source of cold neutrons (CNs) feeding the UCN source. The CN flux in the source can be several times larger than the incident flux, due to multiple neutron reflections from the moderator-reflector. We show that such a source at the PIK reactor would provide an order of magnitude larger density and production rate than an analogous source at the ILL reactor.

We estimate parameters of a $^4$He source with solid methane ($CH_4$) or/and liquid deuterium ($D_2$) moderator-reflector. We show that such a source with $CH_4$ moderator-reflector at the PIK reactor would provide the UCN density of $\sim 1 \cdot 10^5$ cm$^{-3}$, and the UCN production rate of $\sim 2 \cdot 10^7$ s$^{-1}$. These values are respectively 1000 and 20 times larger than those for the most intense UCN user source. The UCN density in a source with $D_2$ moderator-reflector would reach the value of $\sim 2 \cdot 10^5$ cm$^{-3}$, and the UCN production rate would be equal $\sim 8 \cdot 10^7$ s$^{-1}$. Installation of such sources in beams of CNs with equal flux would slightly increase the density and production rate.

**Introduction**

A new concept of super-thermal liquid-helium ($^4$He) UCN sources is presented in ref. [1]. Such sources can be installed in beams of thermal neutrons; thus at ILL (Grenoble) or PIK (Gatchina) such a source would provide parameters highly exceeding those of existing UCN sources. The present work develops ref. [1], considers source parameters in more detail, and applies the new concept to the PIK reactor. Note, however, that such a UCN source can be installed at any other thermal neutron source as well.

The idea of $^4$He UCN sources was proposed in 1975 in ref. [2]; it is based on neutron scattering in liquid $^4$He accompanied with exciting phonons with the energy of 1.02 meV. If the incident neutron energy is slightly higher than 1.02 meV then the cold neutron (CN) is converted into a UCN. As the UCN energy is lower than ~300 neV, only CNs from a very narrow energy range contribute to the UCN production. Cross-sections of simultaneously exciting two or more phonons are lower by a few orders of magnitude. However, the energy of excited phonons is found in a broad range, thus UCNs are produced via multi-phonon processes from a broad spectrum of incident neutrons. Therefore total contributions of one-phonon and multi-phonon process are comparable if the incident neutron spectrum is broad.

Work [2] showed also that produced UCNs can live for a long time in superfluid $^4$He if its temperature is below 1K. Long lifetimes of UCNs allow accumulating them up to high densities. Neutron storage time changes sharply as a function of helium temperature. It equals to the neutron lifetime ~880 s at the temperature of 0.8 K, it is 10 times longer than that at the temperature of 0.6 K, and it is 10 times shorter than that at the temperature of 1 K. Thus it has no sense to decrease the $^4$He temperature below 0.6 K; on the other hand, UCN densities in the source could not reach high values at temperatures above 1 K. The UCN production rate depends on temperature only slightly.

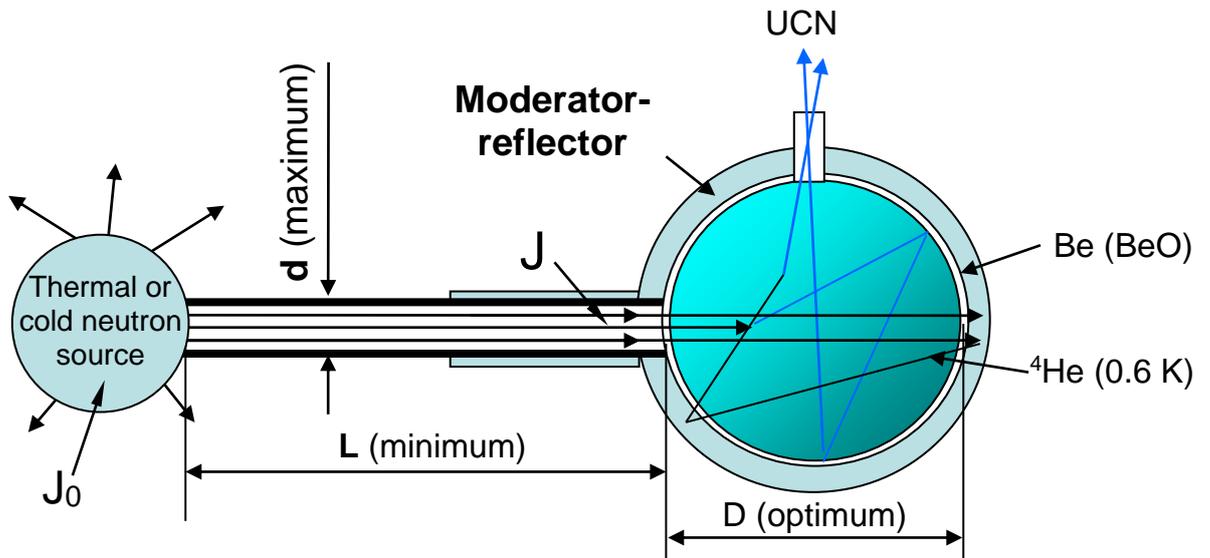

*Fig. 1a. A scheme of UCN source surrounded with moderator-reflector.*

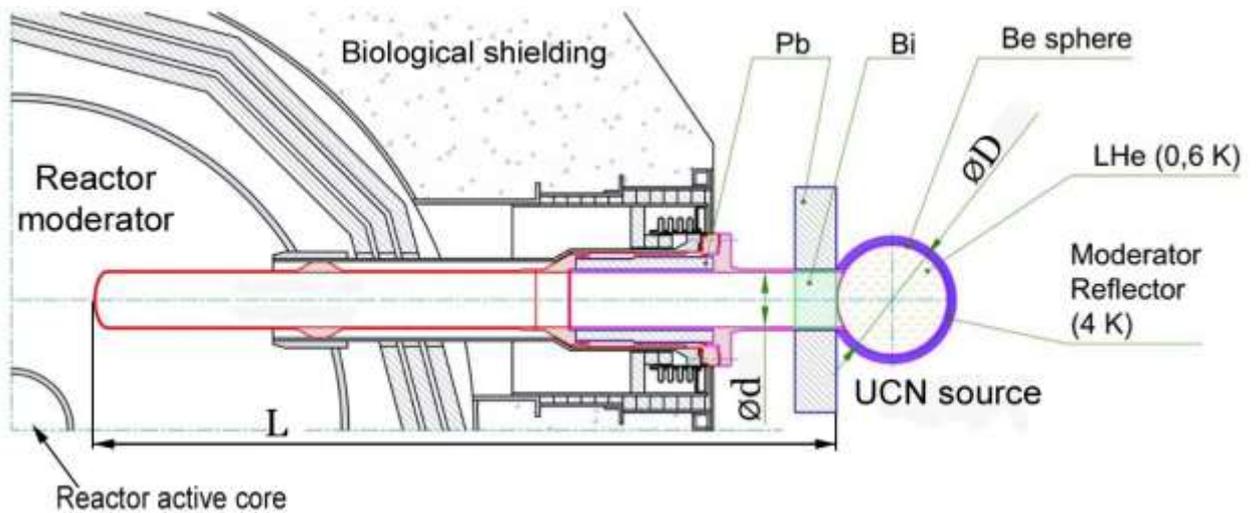

*Fig. 1b. A scheme of possible installation of the UCN source in the beam of thermal neutrons at the PIK reactor.*

There are several operational $^4$He UCN sources in the world: in particular those at KEK (Japan) [3] and ILL [4, 5]. The UCN source at KEK is fed by a neutron source installed at a proton accelerator. UCN sources at ILL are fed by monochromatic neutrons with the wavelength of 8.9 Å, selected from an incident beam of CNs using special crystals-monochromators [16]; one source is used for tests, another one feeds the GRANIT spectrometer [6]. All these sources provide UCN density comparable to that at the UCN facility PF2 at ILL [7]. The production rate in $^4$He sources is much lower than the production rate at PF2.

The new concept [1] consists of installing a $^4$He UCN source in a beam of thermal or cold neutrons at the edge of the biological shielding and surrounding the source with a moderator-reflector (see Fig. 1). The moderator-reflector plays the role of a CN source feeding the UCN source; the CN flux in the UCN source can exceed the incident neutron flux by several times due to neutron multiple reflections.

Such a $^4$He source is a closed spherical trap made from material with high neutron-optical potential (for instance Be) filled in with superfluid $^4$He at the temperature of ~ 0.6 K. Thermal and cold neutrons can easily penetrate through the trap walls, but produced UCNs are trapped; they can be extracted to external experimental setups via a small hole in its upper side.

The UCN production rate is defined by the integral flux of incident neutrons-"parents" as well as by their total path in $^4$He. Thus the shorter and larger guides, the larger UCN production rate and density. For an isotropic source of thermal or cold neutrons with the flux density $J_0$, the flux density J of neutrons incident to the source is proportional to the reciprocal square of the neutron guide length L and to square of the neutron guide diameter d. Thus the integral flux F of incident neutrons is proportional to $d^4$.

The proposed geometry allows profiting from neutrons with a broad angular distribution at the neutron guide exit; otherwise, such neutrons could not be efficiently transported further away from the reactor active zone through mirror guides.

The path of "parents" in $^4$He is proportional to the trap diameter D, to the number of their reflections from the moderator-reflector; also it depends on the probability of their back escape through the entrance neutron guide. Thus the larger trap diameter D, the larger UCN production rate.

Complementary to the production rate, another important parameter is the UCN density accumulated in the source; it is proportional to the production rate and the reciprocal source volume, i.e. to $D^2$.

Thus the source diameter should be optimized in order to balance the two parameters.

Table 1 compares relevant characteristics of PIK and ILL reactors. $L_{min}$ is the minimum neutron guide length defined by the thickness of the reactor biological shielding and the size of the heavy water ($D_2O$) reflector around the active zone; d is the maximum neutron guide diameter defined by the reactor channel size; $J_0$ is the maximum thermal neutron flux density in the $D_2O$ reflector in the vicinity of the reactor active zone; J is the thermal neutron flux density at the guide exit; F is the integral flux of thermal neutrons at the guide exit.

"Optimistic" characteristics are defined by the maximum neutron guide diameter that could be in principle installed. "Realistic" characteristics are defined by the diameter of existing available neutron guides.

| Reactor characteristics | ILL reactor "Optimistic"/"Realistic" | PIK reactor "Optimistic"/"Realistic" |
|---|---|---|
| $L_{min}$ , m | 5 / 5 | 3 / 3 |
| d, cm | 20 / 15 | 30 / 20 |
| $J_0$, $s^{-1}cm^{-2}$ | $\sim 1 \cdot 10^{15}$ / $\sim 1 \cdot 10^{15}$ | $\sim 1 \cdot 10^{15}$ / $\sim 1 \cdot 10^{15}$ |
| J, $s^{-1}cm^{-2}$ | $\sim 1 \cdot 10^{11}$ / $\sim 6 \cdot 10^{10}$ | $\sim 8 \cdot 10^{11}$ / $\sim 4 \cdot 10^{11}$ |
| F, $s^{-1}$ | $\sim 4 \cdot 10^{13}$ / $\sim 1 \cdot 10^{13}$ | $\sim 6 \cdot 10^{14}$ / $\sim 1 \cdot 10^{14}$ |

*Table 1. Comparison of characteristics of PIK and ILL reactors.*

As clear from Table 1, the incident integral neutron flux at the PIK reactor can be an order of magnitude higher than that at the ILL reactor. Respectively, the density and production rate can be an order of magnitude higher at the PIK reactor as well.

Precise calculations for the channel HEC-4 at the PIK reactor (the diameter of 20 cm) using MCNP program resulted to the following estimation of the thermal neutron flux density at the guide exit, three meters away from the reactor active zone center:

$$J=3.66 \cdot 10^{11} \ cm^{-2}s^{-1}, \quad (1)$$

and thus

$$F=1.15 \cdot 10^{14} \ s^{-1} . \quad (2)$$

**UCN source with methane moderator**

We showed in ref. [1] that solid methane ($CH_4$) in phase II at the temperature of $\sim$ 4K is among best materials for moderators-reflectors for such UCN sources.

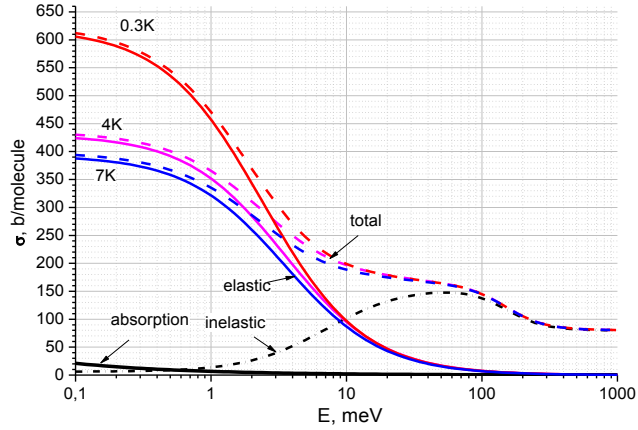

*Fig. 2. The cross section of neutron scattering (per molecule) for solid $CH_4$ in phase II for different temperatures.*

*Solid color lines correspond to cross-sections of elastic scattering at different temperatures. Black dash-dotted line shows the cross section of inelastic scattering. Color dashed lines indicate the total scattering cross section at respective temperatures. Black solid line stands for the absorption cross section.*

Fig. 2 shows cross sections of neutron scattering on solid $CH_4$ as a function of neutron energy at different temperatures (from [8, 9]). As clear from the figure, solid $CH_4$ provides large cross sections of inelastic and elastic scattering; the cross section of elastic scattering increases when the neutron energy decreases. The later property is very important, as it allows accumulating large quantities of CNs inside solid $CH_4$ cavities.

We simulated neutron spectra accumulated in solid $CH_4$ cavities. The calculation was performed using the program MCNP 4c with a special kernel for solid $CH_4$ used in works [9, 10] and kindly provided by the authors.

This calculation yields optimum moderator-reflector parameters. We found that the optimum $CH_4$ thickness is ~2–3 cm; its further increase does not result to the increase in the number of neutrons in the cavity. The optimum $CH_4$ temperature is ~4 K; decreasing it does not help as well. For the neutron guide diameter of 20 cm, the optimum cavity diameter is 40–50 cm; increasing it would significantly decrease UCN density.

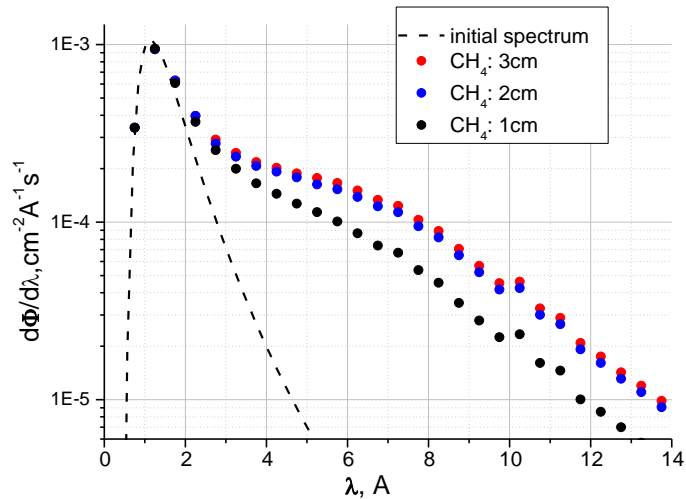

*Fig. 3. Mean fluence, per neutron, in spherical cavity with the diameter of 40 cm surrounded with solid $CH_4$ at the temperature of 4 K.*

*Line corresponds to the mean fluence of incident neutrons (300 K) normalized per one neutron. Points show the mean fluence of accumulated neutrons normalized per one incident neutron for different thicknesses of the cavity walls: 1, 2 and 3 cm.*

Figs. 3, 4 show results of simulations in terms of spectra of neutrons accumulated in a spherical cavity with the internal diameter of 40 cm surrounded with solid $CH_4$. Here $\Phi$ is the fluence per neutron averaged over the cavity volume. $\Phi$ is equal to the ratio of the neutron path in the cavity (till its loss or escape) to the cavity volume. The temperature of initial neutron Maxwell distribution is 300 K; neutrons enter the cavity through neutron guide with the diameter of $\phi$ 20 cm. The fluence of incident neutrons is defined by the neutron path in the cavity until its first interaction with the wall.

To verify calculations presented above, we measured spectrum of neutrons accumulated in a cavity in solid $CH_4$. A comparison of results of this experiment with simulations proved high accuracy of calculations. These measurements are reported in detail in ref. [1].

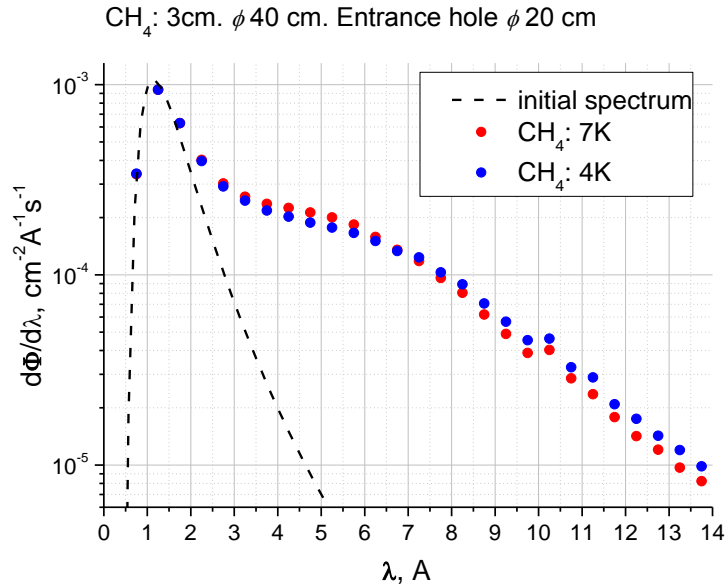

*Fig. 4. Mean fluence, per neutron, in spherical cavity with the diameter of 40 cm surrounded with solid $CH_4$ with the thickness of 3 cm.*

*Line indicates the mean fluence of incident neutrons (300 K) normalized per one neutron. Points correspond to the mean fluence of accumulated neutrons normalized per one incident neutron for different $CH_4$ temperatures: 7 K and 4 K.*

Now we preliminary estimate parameters of a UCN source with the diameter of 40 cm for the integral flux (2) of incident thermal neutrons. At this stage, we do not take into account the influence of liquid $^4$He to neutron fluence (it is not significant) and also the influence of structural materials of the source; these details will be taken into account below.

For one-phonon processes and for UCNs below the beryllium (Be) optical potential (250 neV), the production rate in liquid $^4$He [2] equals:

$R_1 = 4.55 \cdot 10^{-8} \, dJ/d\lambda \, (8.9 \, Å) \, cm^{-3} \, s^{-1}$. (3)

For the incident integral flux $F = 1.15 \cdot 10^{14} s^{-1}$ and the mean fluence per neutron equal $d\Phi/d\lambda \, (8.9 \, Å) = 6.7 \cdot 10^{-5} \, cm^{-2} Å^{-1}$ (see Figs. 3, 4):

$dJ/d\lambda \, (8.9 \, Å) = d\Phi/d\lambda \, (8.9 \, Å) \cdot F = 7.7 \cdot 10^9 cm^{-2} s^{-1} Å^{-1}$ and $R_1 = 350 \, cm^{-3} s^{-1}$. (4)

Estimations of UCN production via multi-phonon processes differ in different works. Thus for neutron spectra (Figs. 3, 4), multi-phonon processes can provide a gain factor in the UCN production rate from 1.25 to 3.5 [11, 12]. We use the lower value: $R = R_1 \cdot 1.25 = 437 \, cm^{-3} s^{-1}$.

The total UCN production rate:

$$P_{UCN} = R \cdot V = 1.5 \cdot 10^7 \, s^{-1}, \quad (5)$$

where V is the source volume.

For the coefficient of UCN loss in the source walls $\eta=10^{-4}$, the mean partial time of UCN storage is $\tau_w \approx 450$ s; taking into account the neutron β-decay it is $\tau \approx 300$ s. For the $^4$He temperature of 0.6 K, UCN partial storage time is 10 times longer than the neutron β-decay lifetime, thus the β-decay could be neglected. The maximum UCN density in the source is:

$$\rho_{UCN}=R \cdot \tau=1.3 \cdot 10^5 \ cm^{-3} . \qquad (6)$$

Now we compare the estimated parameters (5), (6) with those of existing UCN sources.

The most intense user UCN source operates at the ILL [7]. It is a rotating turbine with recessive blades, which decrease the velocity of incident very cold neutrons by their Doppler shifting; a significant fraction of UCNs is lost in the interaction with the blades. The production rate is thus estimated using the integral UCN flux at the source exit. The flux of UCNs with the energy below *250 neV* in the exit guide is $J \sim 3 \cdot 10^4 cm^{-2} s^{-1}$ that corresponds to the density $\rho_{UCN} \sim 10^2 \ cm^{-3}$. The exit neutron guide cross section is *50 cm$^2$*, thus the total UCN production rate is $P_{UCN}=1.5 \cdot 10^6 s^{-1}$.

Thus, the preliminary estimation of the production rate (5) exceeds the production rate in the ILL source by a factor of ~10, while the preliminary estimation of UCN maximum density (6) exceeds the existing density by a factor of ~1000.

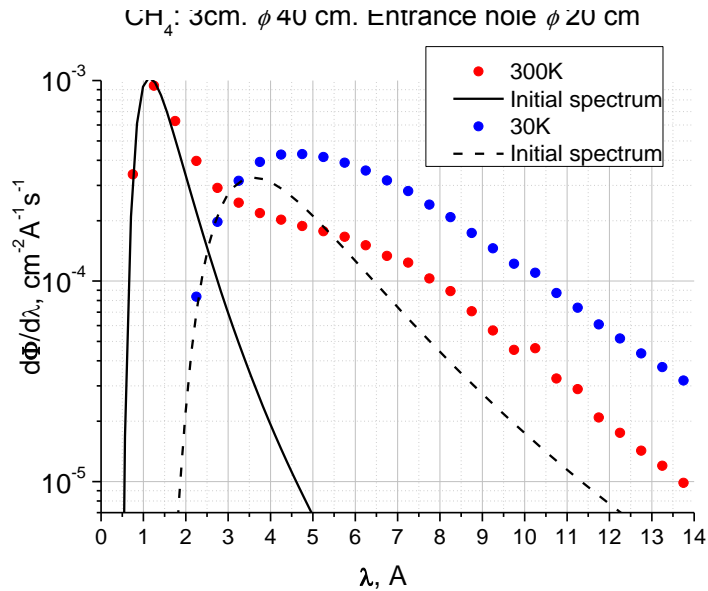

*Fig. 5. Mean fluence, per neuron, in spherical cavity with the diameter of 40 cm surrounded with solid CH$_4$ with the thickness of 3 cm for different spectra of incident neutrons.*

*Lines indicate the mean fluence of incident neutrons (300 K and 30 K) normalized per one neutron. Points show the mean fluence of accumulated neutrons normalized per one incident neutron.*

These calculations were performed for incident thermal neutrons. Fig. 5 compares spectra of accumulated neutrons for thermal (300 K) and cold (30 K) incident neutrons. It follows from Fig. 5 that the use of cold incident neutrons increases the flux of 8.9 Å neutrons and 6 Å neutrons (main "parents" for UCNs produced via multi-phonon processes) by a factor of 2.5 provided that thermal neutrons are totally thermalized and neutron losses in structural materials are absent. Thus parameters (5) and (6) of the UCN source would increase by the same factor. However, real CN sources do not provide total thermalization and are not free of losses in structural materials.

Fig. 6 shows spectra of neutrons incident to the spherical cavity. Curves 1, 2 in this figure correspond to Maxwell spectra with the temperature of 300 K and 30 K, as in Fig. 5. Curve 3 corresponds to the neutron spectrum from liquid-deuterium (D$_2$) CN source with the volume of 14 l placed inside neutron channel in the reactor active zone vicinity. Curve 4 stands for a hypothetical CN liquid-$D_2$ source with the volume of 50 l; a source with so large volume cannot be installed in the reactor active zone vicinity because of the large heat load; all realistic sources would provide spectra somewhere between lines 3 and 4.

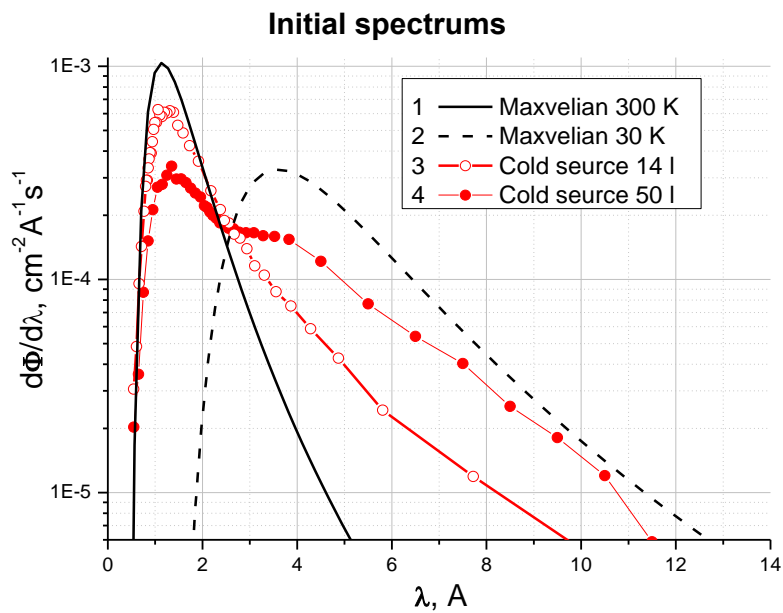

*Fig. 6. Mean neutron fluence incident to spherical cavity with the diameter of 40 cm normalized per one neutron.*
*Lines indicate Maxwell spectra with the temperature of 300 K and 30 K, as in Fig. 5. Open circles correspond to the spectrum of a CN source with the volume of 14 l. Black points show the spectrum from a CN source with the volume of 50 l.*

The integral under any curve in Fig. 6 is the same, because the path in vacuum cavity is equal for all incident neutrons and the curves are normalized per one neutron. Therefore different curves in this figure characterize only the degree of thermalization of neutrons in the CN source. With a certain ambiguity, one could say that a source with the volume of 14 l thermalizes ~20% of neutrons, and a source with the volume of 50 l thermalizers ~50% of neutrons. Calculations show that neutron losses in each source will account for ~30%. Thus a relative increase of CN fluxes is balanced by a decrease of the total flux. Therefore a real CN source would not provide a significant gain in the performance for a UCN source, compared to the configuration with incident thermal neutrons.

**Radiation heating of UCN source**
As mentioned above, to achieve high UCN density in the source, one has to decrease the temperature of liquid $^4$He to ~0.6 K. It is easy to provide such temperature if the heat load is below 1 W; it is difficult to do that if it exceeds 2 W. Thus the heat load should be kept in the range 1–2 W to provide high UCN density. The main heat load arrives from the reactor with radiation.
The radiation heat of the UCN source can be subdivided into two parts:
1. External radiation caused by fast neutrons and γ-quanta arriving to the UCN source from the reactor through the entrance neutron guide.
2. Internal radiation caused by absorption of thermal and cold neutrons ("parents") in construction materials of the UCN source.

**External radiation heat load**

Calculated heat load from the reactor to $^4$He is 3.7 W for the source with the diameter of 40 cm installed in channel HEC-4 of the PIK reactor with the neutron guide diameter of 20 см. This value exceeds considerably our estimated limit of 1–2 W. More than 90 % of the heat load is caused by γ-quanta. Therefore a γ-filter should be installed into the incident neutron beam to scatter and absorb γ-quanta but to transmit neutrons.

Most efficient γ-filters are made of bismuth (Bi). The atomic number of Bi is large thus it interacts strongly with γ-quanta, and its neutron absorption cross section is small. Calculations show that a Bi filter with the thickness of 10 cm decreases the heat load from the reactor down to a value lower than 0.2 W.

As the transmission of thermal neutrons through a polycrystalline filter with the thickness of 10 cm is < 20 %, monocrystalline filters should be used. As the transmission of monocrystalline bismuth depends on crystal mosaicity and its temperature (see [14]), it should be of high quality and kept at low temperature. Thus, a Bi monocrystal at the liquid nitrogen temperature transmits ~60 % of thermal neutrons and >80 % of CNs.

Therefore parameters (5), (6), estimated for incident thermal neutrons should be multiplied by a factor of 0.6 if such a filter is used.

**Internal radiation heat load**

As mentioned above, internal radiation heat load is caused by the neutron absorption in structural materials of the UCN source. For calculating corresponding values, we should consider in more detail the source construction. Fig. 7 shows the geometry of the UCN source.

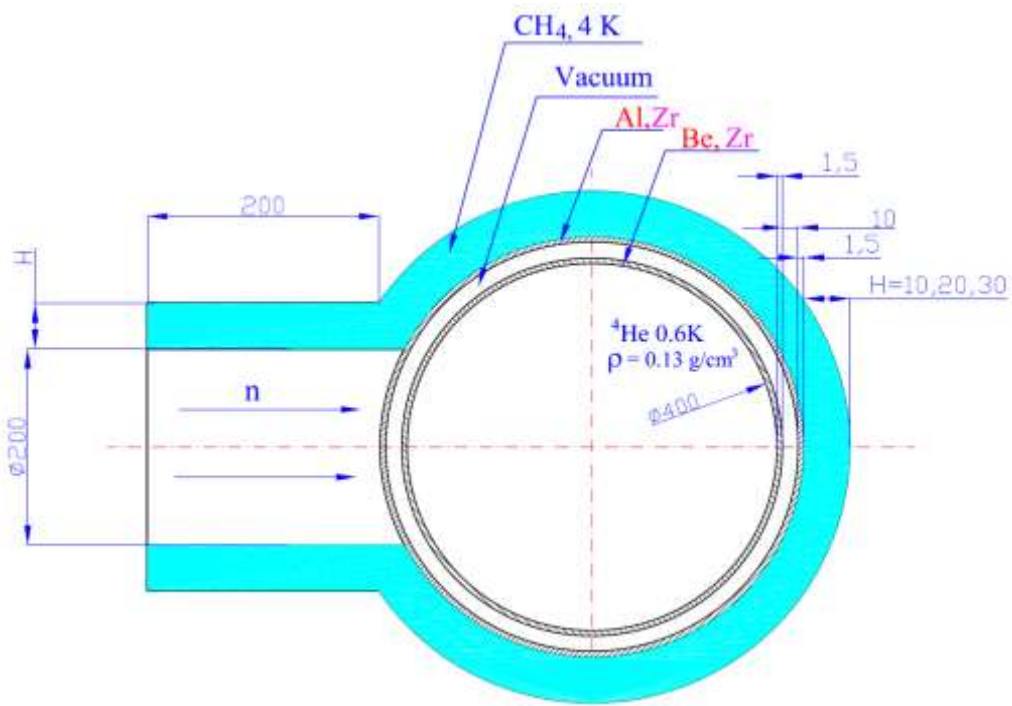

*Fig. 7. A sketch of the UCN source used to calculate the radiation heat load.*

A sketch of the source, shown in Fig. 7, includes two spheres, one inside another one, with the wall thickness of 1.5 mm built of different materials. The internal sphere with the diameter of 40 cm is filled in with liquid $^4$He at the temperature of 0.6 K. The external sphere is separated from it by a vacuum distance of 1 cm.

Materials for the UCN source should meet the following criteria: to provide minimum neutron absorption; to provide minimum heat production that accompanies absorption; to provide minimum interaction with γ-quanta; to provide high neutron-optical potential of the material of internal sphere in order to trap maximum number of UCNs.

Table 2 lists materials appropriate for manufacturing UCN sources:

| Material | Neutron absorption cross section |
|---|---|
| Al | 0.233 b |
| Zr | 0.185 b |
| Be | 0.0076 b |
| C (pyrographite) | 0.0035 b |

*Table 2. The neutron absorption cross section (for thermal neutrons) for materials appropriate for manufacturing UCN sources.*

Among these materials, aluminum (Al) provides best structural properties, however it releases a large amount of heat after neutron capture. Therefore it could be used only for manufacturing the external sphere, which is not in direct contact with liquid $^4$He.

Among these materials, only Be provides high neutron optical potential. If other materials are used for the internal sphere, they should be coated with another material with high neutron optical potential. Table 3 lists neutron optical potentials $E_b$, and respective critical velocities $v_b$, for materials appropriate for coating the internal sphere.

| Material | $E_b$ ($v_b$) |
|---|---|
| Be | 250 neV (6.9 m/s) |
| Diamond | 300 neV (7.6 m/s) |
| DLC (sp$^3$-70%) | 270 neV (7.2 m/s) |
| DLC (sp$^3$-45%) | 250 neV (6.9 m/s) |

*Table 3. Neutron optical potentials (critical velocities) for various coatings of the internal UCN trap.*

DLC in this table is diamond-like material, which contains both diamond fraction (sp$^3$-hybridization), and graphite fraction (sp$^2$-hybridization).

After giving these details, we return to calculations of internal radiation heat load. Fig. 8 shows estimations of the heat load and also the UCN production rate in the source. The calculation is performed for incident thermal neutrons with the flux density $3.66 \cdot 10^{11}$ cm$^{-2}$s$^{-1}$ (the integral flux $1.15 \cdot 10^{14}$ s$^{-1}$) for different thicknesses of solid CH$_4$. Both the heat load in liquid $^4$He and the heat load in the internal sphere material are taken into account.

The main reason for UCN trap heating is γ-quanta emitted after absorption of neutrons in CH$_4$, and then interacting with $^4$He and the internal sphere material. When the CH$_4$ thickness decreases, the UCN production rate decreases less sharply than the heat load does. Thus the optimum CH$_4$ thickness is found in the range 1.5–2 cm. *Be* nearly does not affect the heat load. The zirconium (Zr) atomic weight is larger than that of Be, thus the cross section of its interaction with γ-quanta is higher, and it is more heated.

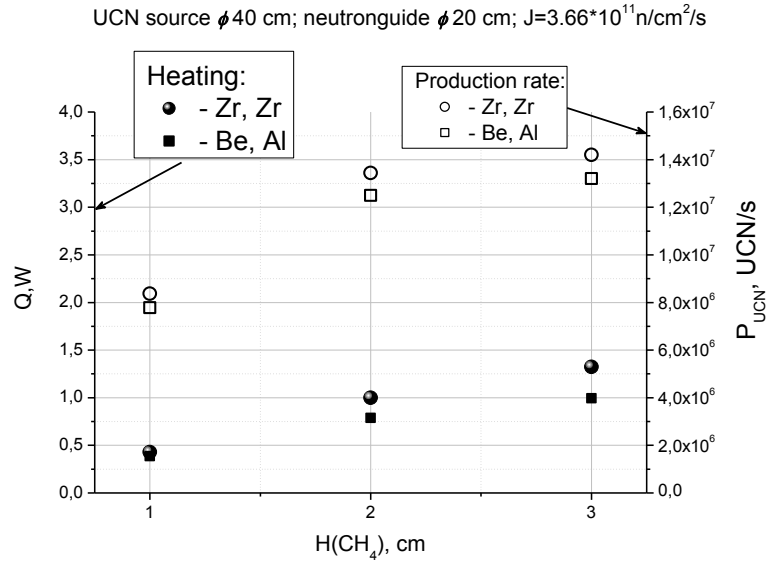

*Fig. 8. Heat load in the UCN trap (see Fig. 7) caused by internal radiation (on left); and UCN production rate (on right).*

*Open points show UCN production rate. Black points indicate heat load. Squares correspond to calculations for external Al and internal Be spheres. Circles correspond to calculations for both Zr spheres.*

Calculation shown in Fig. 8 is performed with no Bi-filter in the incident neutron beam. For a filter with the transmission of 60%, all values should be multiplied by a factor of 0.6.

**Final results for the UCN source with $CH_4$ moderator**

Here we summarize results of performed calculations. Table 4 lists parameters of the UCN source with Be trap with the diameter of D=40 cm, for the diameter of the neutron guide d=20 cm, with a Bi-filter with the transmission for thermal neutrons equal 60%, for the flux density of incident neutrons $J = 3.66 \cdot 10^{11}$ cm$^{-2}$s$^{-1}$, and for the $CH_4$ thickness of $H_{CH4}$=2 cm.

| Source parameters | Thermal neutrons |
|---|---|
| $P_{UCN}$, s$^{-1}$ | $7.5 \cdot 10^6$ |
| $\rho_{UCN}$, cm$^{-3}$ | $6.7 \cdot 10^4$ |
| Q, W | 0.65 |

*Table 4. Parameters of the UCN source (D=40 cm, d=20 cm, $J=3.66 \cdot 10^{11}$ cm$^{-2}$s$^{-1}$, $H_{CH4}$=2 cm and Bi-filter) for incident thermal neutrons.*

*$P_{UCN}$ stands for the UCN production rate; $\rho_{UCN}$ indicates the maximum UCN density in the source, Q is the heat load to the source.*

The design of HEC-4 channel in the reactor PIK allows installing there a neutron guide with the diameter of d=25 cm. In this case, to limit the escape of CNs from the UCN source back to the neutron guide, one should increase proportionally the source diameter up to D=50 cm. This modification increases the incident neutron flux density proportionally to d$^2$, up to $J = 5.72 \cdot 10^{11}$ cm$^{-2}$s$^{-1}$, and the production rate increases proportionally to d$^5$, i.e. by a factor of 3; the external heat load increases proportionally as well. Parameters of such a source are listed in Table 5.

| UCN source parameters | Thermal neutrons |
|---|---|
| $P_{UCN}$, s$^{-1}$ | $2.2 \cdot 10^7$ |
| $\rho_{UCN}$, cm$^{-3}$ | $1.0 \cdot 10^5$ |

| Q, W | 1.95 |

*Table 5. Parameters of the UCN source (D=50 cm, d=25 cm, J=5.72·10¹¹ cm⁻²s⁻¹, $H_{CH_4}$=2 cm and Bi-filter) are given for incident thermal neutrons.*

As clear from Table 5, we reach the upper limit for the heat load equal 2 W. The heat load can be decreased by increasing the Bi-filter thickness or/and by decreasing the $CH_4$ thickness, however, anyway it is close to the limit.

The calculated production rate exceeds the production rate in the ILL UCN source by a factor of 20; the calculated maximum UCN density exceeds the ILL UCN density by a factor of 1000. These values are not preliminary estimations (5), (6), but the calculations performed for a realistic geometry shown in Fig. 7.

A question arises: "Can we improve even further the parameters of UCN source of this type and simultaneously decrease the heat load?" The answer to this question is: "Yes, we can. We should use another moderator".

**UCN source with $D_2$ moderator**

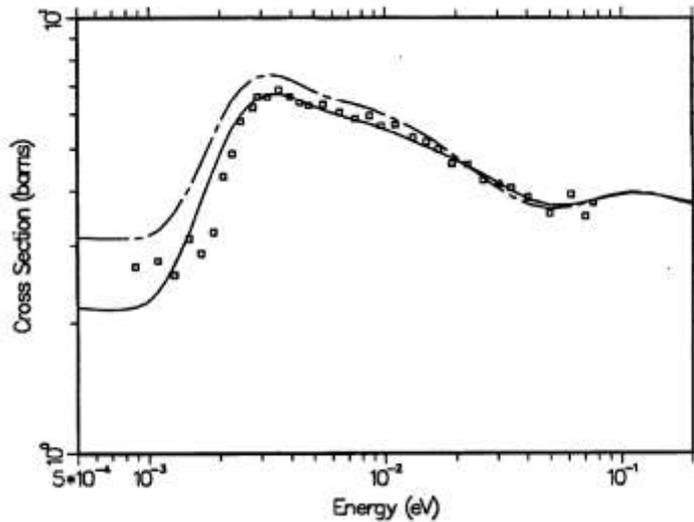

*Fig. 9. Total cross section of neutron scattering (per atom) on liquid $D_2$ at the temperature of 19 K. Upper curve corresponds to orto-$D_2$, lower curve – to para-$D_2$, and points show results for equilibrium orto-para mixture at the temperature of 19 K.*

Liquid $D_2$ is usually used as moderator for producing CNs. This fact is due to the record ratio of the scattering cross section to the absorption cross section for this material. The absolute cross section of neutron scattering on $D_2$ is lower than that on $CH_4$ by two orders of magnitude, however the absorption cross section is smaller by three orders of magnitude than that for $CH_4$. However, in order to profit from this advantage and efficiently moderate neutrons, one has to provide large $D_2$ thickness.

Fig. 9 shows the total cross section of neutron scattering on liquid $D_2$ at the temperature of 19 K [15].

Figs. 2, 9 indicate that the cross section of scattering on $D_2$ decreases at smaller neutron energies, while it increases at smaller neutron energies for $CH_4$. We will use this feature.

Fig. 10 compares mean fluence of CNs accumulated in the same cavity as considered above (D=40 cm, d=20 cm) for $CH_4$ moderator with the thickness of 3 cm and for $D_2$ moderator with the thickness of 50 cm. As clear from the figure, the fluence of 8.9 Å neutrons is slightly lower for $D_2$ moderator; the fluence of 6 Å is slightly higher for $CH_4$ moderator. Thus production rates for such UCN sources are close to each other.

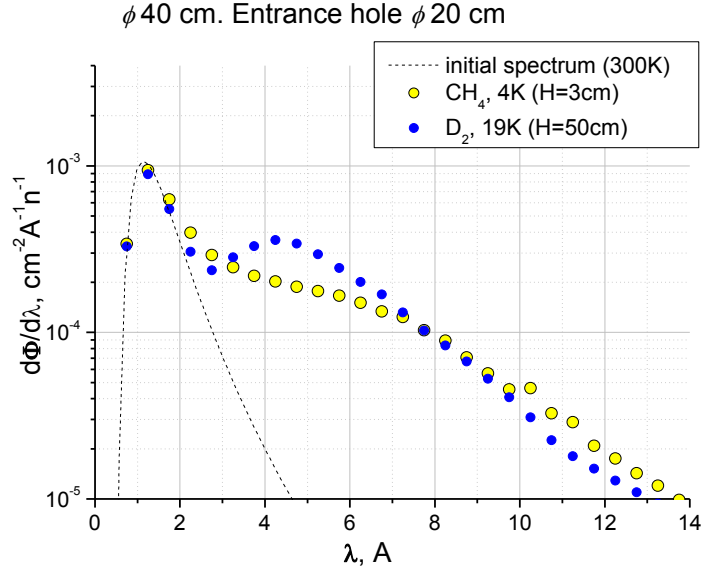

*Fig. 10. Mean fluence per neutron in spherical cavity with the diameter of 40 cm for $CH_4$ and $D_2$ moderators.*

*Curves indicate the mean fluence of incident neutrons (300 K) normalized per one neutron. Points correspond to the mean fluence of accumulated neutrons, normalized per one neutron, for $CH_4$ with the thickness of 3 cm at the temperature of 4 K as well as for liquid $D_2$ at the temperature of 19 K with the thickness of 50 cm.*

Consider increasing the source diameter D. Fig. 11 shows the UCN production rate and the maximum UCN density as a function of the trap diameter; the incident flux of thermal neutrons (T=300 K) is always equal $F=1.15 \cdot 10^{14}$ $s^{-1}$. Calculation, presented in this figure, is performed in simplified geometry and approximations as those used for estimations (5), (6).

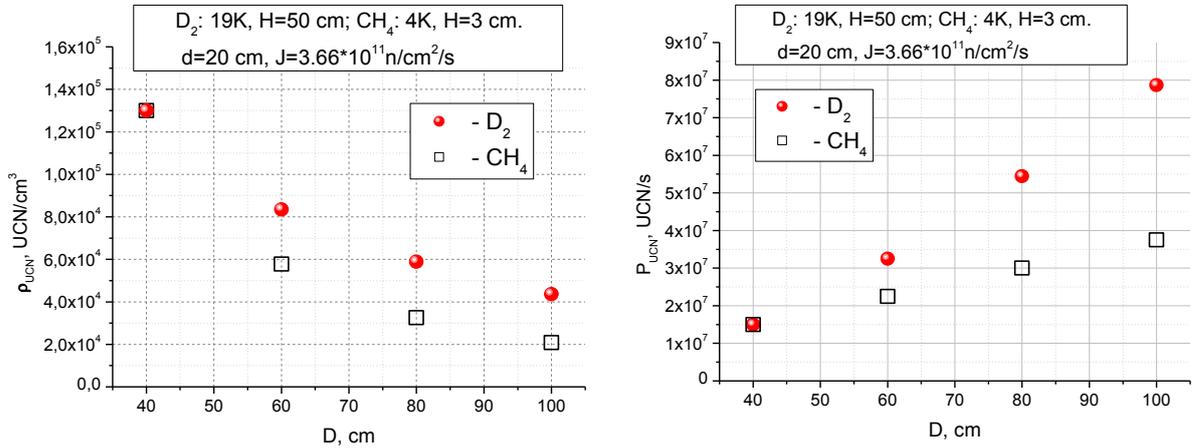

*Fig. 11. UCN production rate (on right) and maximum UCN density (on left) as a function of the source diameter.*

*Squares correspond to $CH_4$ moderator with the thickness of 3 cm. Circles indicate results for $D_2$ moderator with the thickness of 50 cm.*

As clear from Fig. 11, the production rate in $D_2$ source increases much sharper than that in $CH_4$ source. This difference is due to the fact that neutron albedo from $CH_4$ stays the same as the $CH_4$ thickness is much smaller than the source diameter, while neutron albedo from $D_2$ increases as its thickness is comparable to the source diameter; neutrons can find the cavity with $^4He$ easier if cavity is large. Therefore, in order to favor $D_2$ moderator, one has to increase the source diameter D.

However, a source with the moderator thickness of 50 cm contains as much as 1.5 m$^3$ of liquid $D_2$ even for the source diameter D=40 cm. Working with so large quantity of $D_2$ is expensive, technically complicated and dangerous. As mentioned above, the cross section of scattering of neutrons on liquid $D_2$ drops down considerably at the energy of ~1 meV, i.e. precisely for neutrons with the wavelength of 8.9 Å. If liquid $D_2$ is surrounded by a layer of solid $CH_4$, such a combination decreases considerably the quantity of $D_2$, without affecting considerably the UCN production rate.

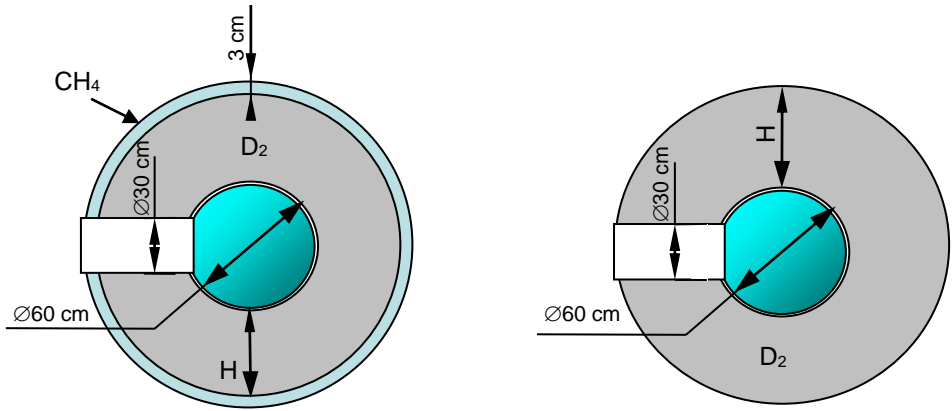

*Fig. 12. A sketch for calculating $D_2$ source surrounded with a solid $CH_4$ layer (on left), and that with no $CH_4$ (on right).*

The following calculations are carried out for the source with the diameter of D=60 cm and the entrance neutron guide with the diameter of 30 cm (the maximum possible diameter for a neutron guide at the PIK reactor). The incident neutron flux density is then equal $J=8.2 \cdot 10^{11}$ cm$^{-2}$s$^{-1}$ and the integral flux is $F=5.8 \cdot 10^{14}$ s$^{-1}$. A source sketch for this calculation is shown in Fig. 12.

Results of this calculation are shown in Figs. 13, 14 for incident thermal neutrons (Maxwell distribution with the temperature of 300 K).

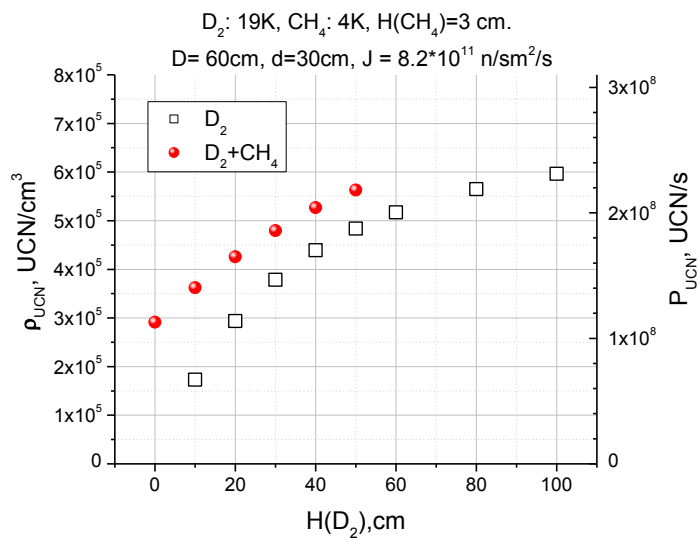

*Fig. 13. Production rate (on right) and maximum UCN density (on left) as a function of $D_2$ thickness for incident thermal neutrons (300 K), with no Bi-filter.*

*Squares correspond to moderator with no $CH_4$. Circles indicate results for moderator with a $CH_4$ layer with the thickness of 3 cm.*

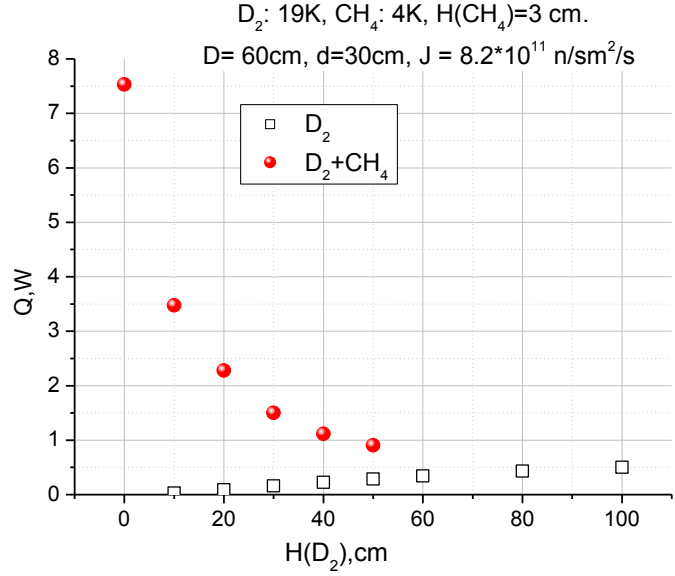

*Fig. 14. Internal radiation heat load in the $^4$He source as a function of $D_2$ thickness for incident thermal neutrons (300 K) with no Bi-filter.*
*Squares correspond to calculations with no $CH_4$. Circles indicate results of calculations with a layer of $CH_4$ with the thickness of 3 cm.*

As clear from Fig. 13, a thin layer of $CH_4$ on the surface of $D_2$ moderator with the diameter of 60 cm replaces 20 cm of $D_2$. Increasing the thickness of $D_2$ in this combined moderator sharply decreases the heat load, because the main source of internal radiation is γ-quanta emitted from $CH_4$, thus increasing the $D_2$ thickness we replace the source of γ-quanta further from $^4$He.

For a UCN source with purely $D_2$ moderator, the dominant heat load is provided by external radiation heating, which results in ~1.5 W, provided the Bi-filter thickness is 10 cm and the neutron guide diameter is 30 cm.

Table 6 presents parameters of the UCN source with Be trap with the diameter of D=60 cm, with the neutron guide diameter of d=30 cm, with Bi-filter in the incident beam (the transmission for thermal neutrons is 60%), with $D_2$ moderator with the thickness of 50 cm, for the incident neutron flux F=5.8·10$^{14}$ s$^{-1}$.

| UCN source parameters | Thermal neutrons |
|---|---|
| $P_{UCN}$, s$^{-1}$ | 8.3·10$^7$ |
| $\rho_{UCN}$, cm$^{-3}$ | 2.2·10$^5$ |
| Q, W | 1.7 |

*Table 6. Parameters of the UCN source with $D_2$ moderator (D=60 cm, d=30 cm, J=8.2·10$^{11}$ cm$^{-2}$s$^{-1}$, $H_{D2}$=50 cm with Bi-filter in the incidence beam) for incidence thermal neutrons.*

**Extraction of UCNs to external setups**

Above, we analyzed UCN sources in terms of UCN production rate and UCN density in the source. For practical applications, however, two more parameters are important.

First, we consider extraction of UCNs from the source to external setups. In fact, UCN flux to and UCN density in an external experimental setup are important. These values depend on geometry of the source, the setup and the extraction lines. Such configurations might be quite different. UCN density in a small setup could reach nearly the maximum UCN density in the source, however the detection rate is always much smaller than the production rate. In a setup with a volume, say, an order of magnitude larger than the source volume, UCN density is "diluted", it is an order of magnitude lower than the maximum UCN density but the detection rate could approach the UCN production rate. In a setup equivalent to "black body" connected via efficient extraction system, UCN detection rate is equal to the UCN production rate. Thus both UCN density and UCN production rate are important for planning experiments.

For setups of each type, an optimum method of UCN extraction should be applied. Thus, for large installations with significant UCN losses, a maximum number of UCNs should be delivered to the setup, not necessarily through a large exit window in the source. Thus, for a source with the volume of 65 l (a sphere with the diameter of 50 cm), for the coefficient of UCN loss in the source equal $1\cdot10^{-4}$, ~ 90% of UCNs can be extracted through a hole with the area as small as 10 cm$^2$. If an expanding mirror conus is installed behind the source exit window, it would produce UCN beam with an angular distribution extended along its axis; such a beam can be transported through a mirror neutron guide with the diameter of 6-7 cm nearly loss-free to distances of up to ~10 m that is usually sufficient. At the entrance to the setup, the beam can be refocused to a square area of ~10 cm$^2$ if needed. The window at the source exit with the square area of 10 cm$^2$ decreases the UCN density in the source, and thus at the exit from the source, by a factor of 10. For a small setup requiring maximum density, the window at the exit from the source should be decreased down to ~1 cm$^2$. Feasibility of such a universal system of extraction and transport requires further detailed studies.

Second, one should note that the task of building a UCN production trap from a technologically convenient material with small cross section of neutron loss (the loss coefficient ~$1\cdot10^{-4}$) and high optical potential (>200 neV) is not yet solved and requires a dedicated study.

**Conclusion**

The goal of this article is to explore general features of production of UCNs in $^4$He sources of a new type, as well as to optimize parameters of such a source.

Presented calculations show that a $^4$He UCN source with a moderator-reflector from solid CH$_4$, installed in a beam of thermal neutrons at the PIK reactor, would provide the UCN density of ~$1\cdot10^5$ cm$^{-3}$ in the source and the UCN production rate of ~$2\cdot10^7$ s$^{-1}$. These values are much larger than parameters of existing UCN sources.

The UCN density in a source with D$_2$ moderator can reach the value of ~$2\cdot10^5$ cm$^{-3}$, while the production rate can be as high as ~$8\cdot10^7$ s$^{-1}$.